# Description of a specialized stress equipment for EPR X-band measurements

Th.W. Kool
University of Amsterdam

**Abstract**
A specialized stress equipment has been developed for EPR X-band experiments.
Uniaxial stress experiments in EPR of different impurity ions in $SrTiO_3$ and $BaTiO_3$ will be reviewed.

**Introduction**

Stress measurements in electron paramagnetic resonance (EPR) X-band is notably difficult to perform because of lack of space in rectangular and cylindrical resonant cavities. By putting stress equipment in the cavity, often consisting of steel or brass, results in a considerable lowering of the quality factor $Q$ expressing the ratio between its energy content and its energy loss per period [1]. Furthermore perovskite crystals like $SrTiO_3$ and $BaTiO_3$ with their high dielectric constants, especially at low temperatures, lower the $Q$ value considerably and hence the sensitivity of the measurements. See for values of the dielectric constant ref. 2 for $SrTiO_3$ and ref. 3 for $BaTiO_3$. In EPR K-band measurements (with a resonant cavity much smaller in size than in X-band) Berlinger and Müller solved this problem by putting the cavity and stress equipment in a helium Dewar immersed in liquid helium or liquid nitrogen. Intermediate temperatures between 4.2 K and room temperature were obtained by heating the cavity walls with a controlled heater. The equipment has been extensively described in an article by Müller and Fayet [4]. In their equipment the stress can be applied perpendicular to the magnetic field. This article has also been published in a book written by Noble laureate Alex Müller and Tom W. Kool [5].

EPR stress experiments are extremely useful in studying structural phase transitions, Jahn-Teller (JT) and off-centre impurities [4, 5].

In this article we will describe stress equipment developed for X-band measurements and review results for different paramagnetic impurities in single crystals of the perovskites $SrTiO_3$ and $BaTiO_3$ measured at the University of Amsterdam and the University of Osnabrück. Some of the results are published in the Müller-Kool book [5].

**Stress equipment**

Due to the experimental set up in X-band measurements the stress equipment has to be mounted into the cavity. This influences the resonant cavity, which was of a rectangular optical transmission cavity of $TE_{001}$ type. The stress apparatus is placed perpendicular to the applied magnetic field. A schematic drawing of the stress equipment is given in the figure on following page. The device is transforming the pressure in a gas chamber into uniaxial stress through a connection with a stainless steel rod. The bottom end of this rod was connected with two pieces of Teflon rods in order to improve the homogeneity of the applied stress. Small cylindrically or rectangular shaped crystals, with size typically 1×1×2 mm$^3$, were inserted between the Teflon rods. Also the surfaces of the crystals were carefully polished in order to optimize contact with the Teflon rods and to prevent cracking of the crystals. The Teflon rod was mounted into a small stainless steel pipe or a quarts pipe. The stainless steel pipe had vertical elongated holes in it to realize the least possible quantity of steel in the cavity resulting in a better $Q$-value.

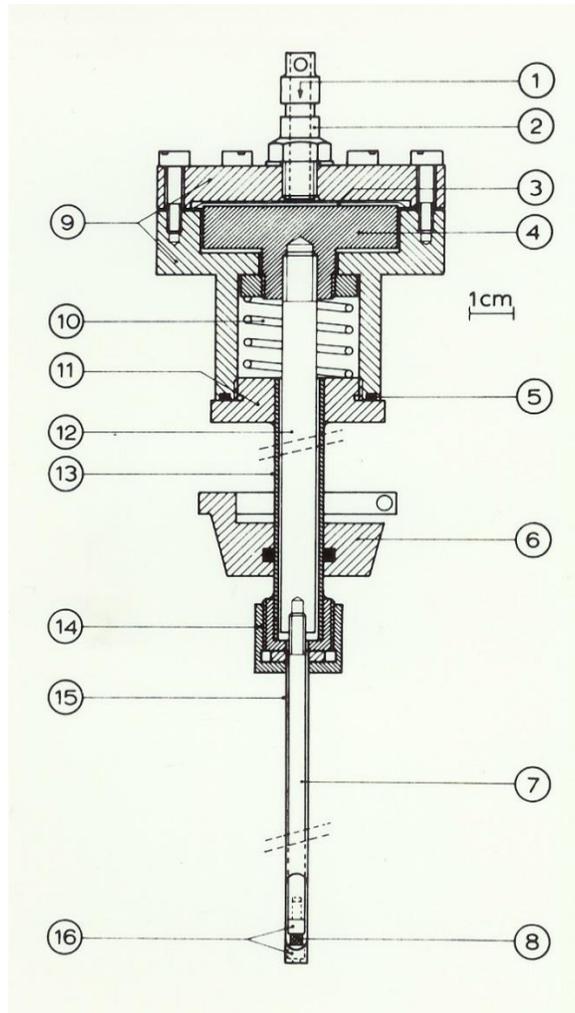

1. Gas inlet
2. Quick-seal connector
3. Silicon rubber membrane
4. Plunger
5. O-ring
6. Conical seal
7. Brass pressure rod
8. Crystal
9. Pressure house
10. Weight compensation spring
11. Spring holder
12. Stainless steel pressure rod or quartz pressure rod
13. Rod guidance
14. Connector
15. Quartz guidance pipe
16. Teflon plates

At liquid helium temperatures we used a quartz pipe, because otherwise due to the high dielectric constants, especially of $SrTiO_3$, we were not able to tune the cavity. At these temperatures additional microwave modes in the sample were observable. As said before it is important to use tiny crystals typically in the order of 2-3 mm$^3$. With the quartz rod we were able to achieve stresses up to $8.5 \times 10^8$ dynes cm$^{-2}$.

**Stress results for different impurity ions in $SrTiO_3$ and $BaTiO_3$**

Stress experiments in EPR are very useful to discriminate between Jahn-Teller (JT) impurities, off-centre systems, hole like ions and impurities with associated defects or vacancies. The first three mentioned type of impurities can be reoriented by stress in contrast to the associated defects impurities. This is observable in EPR by a change in line intensities. The used stress equipment can be applied perpendicular to the magnetic field.

**Impurities in $SrTiO_3$**

Most of the discussed impurity ions are surrounded by an octahedron of six $O^{2-}$ ions. Furthermore the JT-ions are of a tetragonal $T \otimes e_g$ type with the exception of orthorhombic $Cr^{5+}$ which is of a $T \otimes (e_g+t_{2g})$ type JT system.

*$SrTiO_3$: tetragonal $Cr^{5+}$*. In one model the tetragonal $Cr^{5+}$ is located at the centre of an oxygen octahedron (surrounded by six $O^{2-}$ ions). The stress experiments were interpreted as a squeezing of the oxygen octahedron due to a $T \otimes e_g$ type J-T coupling [6].
Another explanation uses as model that the impurity centre sits off-centre in one of the [100] directions in an oxygen cage due to a smaller ion radius (0.54 Å) in comparison with that of the $Ti^{4+}$ ion (0.64 Å) which it replaces [7, 8]. In a future article we will show that the latter model is more reliable [9].
*$SrTiO_3$: orthorhombic $Cr^{5+}$*. Stress experiments showed that the JT-ion is orthorhombically distorted [10, 11]. Later, Müller, Blazey and Kool gave arguments that the ion must sit off-centre in one of the [110] directions tetrahedrally surrounded by four oxygen ions [12]. The ion radius of $Cr^{5+}$ is only 0.34 Å in tetrahedral symmetry, much smaller than that of $Ti^{4+}$ which it replaces [8]. This has been confirmed by static electric field experiments [13].
*$SrTiO_3$:$V^{4+}$*. Stress experiments showed that this JT-ion sits on-centre in in squeezed oxygen octahedron, which was also confirmed by static electric field experiments [14, 15].
*$SrTiO_3$:$Fe^{2+}$-$O^-$*. The axes of this small polaron hole centre could be altered by uniaxial stress experiments. Both stress and static electric field experiments gave rise to a change in direction of the axes of this centre [16]. Also a differential stress coupling coefficient could be derived.
*$SrTiO_3$:$V^{5+}$-$O^-$*. In the thesis of Lagendijk this centre was assigned as a hole centre [17]. Stress experiments by Kool showed that the axes of this centre could not be changed. Therefore the centre must be a linear $Fe^{5+}$-$O^{2-}$-$V^{5+}$ type [18]. With the help of the stress experiments a linear stress tensor could be obtained.

**Impurities in BaTiO$_3$**

The stress equipment used in Osnabrück was almost the same as described before with minor alterations. All the discussed JT-ions are of a T$\otimes$e$_g$ or EType hier uw vergelijking. e$_g$ type. Most of the experiments are described in the thesis of Susanne Lenjer [19].

*BaTiO$_3$: oxygen vacancies.* Ti$^{3+}$-V$_o$ and Ti$^{3+}$-V$_o$ associated with Na or K respectively. These centres could not be altered by stress [20].
*BaTiO$_3$ acceptor-poor samples.* Low concentration of compensating oxygen vacancies V$_o$ contain isolated Ti$^{3+}$ free small polarons. These small polarons can be reoriented under uniaxial stress.
*Acceptor-rich crystals* show Ti$^{3+}$ next to V$_o$ alkali acceptor complexes: Ti$^{3+}$-V$_o$-K$_{Ba}$. The K replaces Ba$^{2+}$. These centres could not be reoriented by stress [21].
*BaTiO$_3$:Mo$^{5+}$.* With the help of stress experiments is was shown that this is a static JT-ion. The stress experiments revealed a reorientation of the centre axes. The ion has almost the same radius as Ti$^{4+}$ which it replaces, therefore it cannot move off-centre and must be a JT-ion [22-23].
*BaTiO$_3$: small polarons and intermediate polarons.* With the help of stress experiments interesting behaviour of these polarons could be studied [22-23].
*BaTiO$_3$:Ni$^+$.* The axes of this JT-centre can be changed by stress. At low temperatures the centre is static. At higher temperatures there is a transition from static to dynamic JT behaviour [24].
*BaTiO$_3$:Rh$^{2+}$.* This centre is also a static JT-centre confirmed by stress experiments [24].
*BaTiO$_3$:Ni$^+$ substituted for Sr$^{2+}$.* In contrary to the other discussed centres this ion is replacing a Sr$^{2+}$ ion and sits off-centre. Stress experiments gave rise to a reorientation of the centre axes [25].
*BaTiO$_3$:Pt$^{3+}$.* Stress experiments confirmed that this is a static JT centre [24].
*BaTiO$_3$:Fe$^{5+}$.* Possenriede *et al* described this impurity as an Fe$^{5+}$ centre with an associated Ti$^{4+}$ vacancy in a neighbouring oxygen octahedron a so-called Fe$^{5+}$-O$^{2-}$-V$_{Ti}$ [25]. Stress experiments showed that Fe$^{5+}$ actually sits off-centre in one of the [111] directions [24]. This is supported by the small ionic radius of the ion.
*BaTiO$_3$:Na$^+$-O$^-$ small polaron.* In these crystals alkali ions Na and K replace Ba forming Na$^+$-O$^-$ and K$^+$-O$^-$ dipoles and can be reoriented under uniaxial stress. Also a differential stress coupling coefficient could be derived.